# SINGULARITY AND MESH DIVERGENCE OF INVISCID ADJOINT SOLUTIONS AT SOLID WALLS

**CARLOS LOZANO[1] AND JORGE PONSIN[2]**

Computational Aerodynamics Group
National Institute of Aerospace Technology (INTA).
Carretera de Ajalvir, km 4. Torrejón de Ardoz 28850, Spain.
[1]lozanorc@inta.es, [2]ponsinj@inta.es

## 1 INTRODUCTION

Recently, it was found that certain adjoint solutions to the Euler flow equations fail to converge with mesh refinement in the proximity of solid walls [1]. The purpose of this paper is to show that this behavior can be explained using an analytic adjoint solution obtained in [2]. The problem is best illustrated with a simple example: the inviscid incompressible flow at angle of attack $\alpha$ = 0º past a symmetrical van de Vooren airfoil with 12% thickness and trailing edge angle $\tau = 16°$ built from the circle $\zeta = Re^{i\theta}$, where $0 \le \theta < 2\pi$, via the conformal transformation [3]

$$z(\zeta) = \frac{(\zeta - R)^k}{(\zeta - \varepsilon R)^{k-1}} + 1 \tag{1}$$

with $R = (1+\varepsilon)^{k-1} / 2^k$, $\varepsilon = 0.0371$ and $k = 86/45$. The trailing edge is at $z$ = 1, which corresponds to $\zeta_{te} = X_{te} + iY_{te} = (R,0)$ in the circle plane.

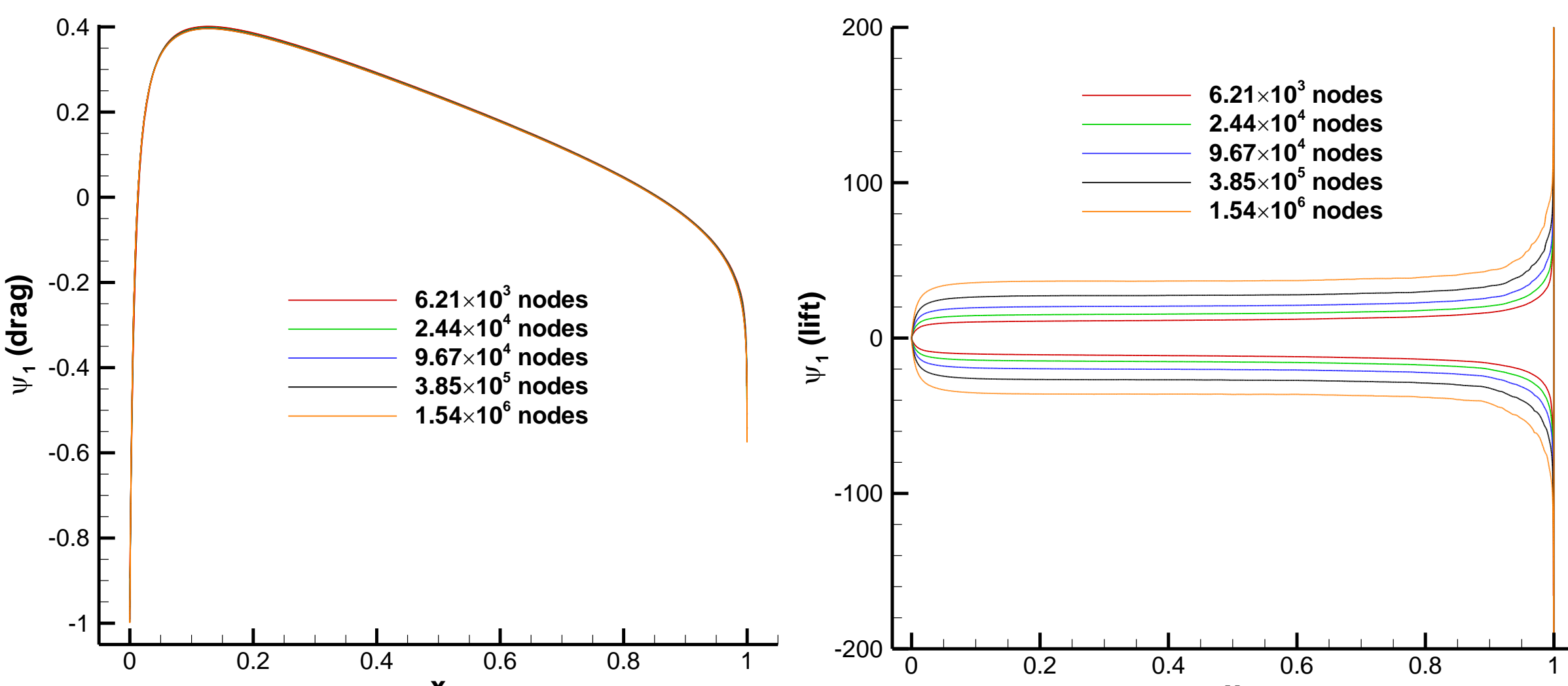

**Figure 1**. Drag (left) and lift (right)-based inviscid incompressible adjoint solution on the van de Vooren airfoil profile (1) at $\alpha$ = 0º computed with the SU2 solver on 5 progressively refined unstructured triangular meshes.

Figure 1 shows the adjoint values on the airfoil profile computed with the SU2 incompressible solver [4] on a series of uniformly refined unstructured triangular meshes. The farfield boundary is placed at 100 chords. The coarsest mesh in the series has 6211 nodes and is shown in Figure 2 (left). Each mesh in the series is constructed from the previous one by subdividing each triangle into 4 self-similar triangles. Boundary edges are made to conform to the airfoil shape by Bézier-spline surface reconstruction. It is observed that the drag-based adjoint solution (left) behaves smoothly and converges with mesh refinement, while the lift-based adjoint solution (right) appears to be singular at the trailing edge on any given mesh and the value along the remainder of the airfoil grows continually as the mesh density increases.

## 2 CHARACTERIZATION OF THE MESH DIVERGENCE PROBLEM

In order to characterize the problem, several tests have been performed [1] [5], with the following conclusions:

1. The problem is limited to inviscid cases.
2. The adjoint-based sensitivity derivatives are not affected.
3. The issue depends on the cost function and flow regime as follows:
   - In supersonic flow, lift or drag-based adjoint solutions do not show this behavior.
   - In transonic and subsonic flow, including incompressible flow, lift-based adjoint solutions are always affected, while drag-based solutions are only affected for transonic rotational flows (e.g. shocked flow past a symmetric airfoil with non-zero angle of attack).
   - The adjoint state based on the far-field entropy flux shows the same behavior as the near-field drag.
4. The problem is not exclusive of a particular solver or numerical scheme, having been observed with wildly different solvers (DLR's Tau code [6], Stanford University's SU2 code [4], ONERA's ELSA code [7] and Imperial College's Nektar++ code [8]).
5. The issue appears in two and three dimensions.
6. The adjoint wall boundary conditions is reasonably well obeyed across mesh levels except in the proximity of the trailing edge.
7. The anomaly is observed in all types of trailing edge configurations (including blunt and cusped trailing edges) and also in blunt bodies such as circles and ellipses.
8. The issue does not seem to depend on the far-field distance, resolution or the adjoint far-field b.c.
9. The anomaly is tied to adjoint singularities at the trailing edge or rear stagnation point and at the incoming stagnation streamline.
10. It was shown that increasing dissipation levels did not prevent the mesh divergence, but the actual value of the adjoint solution at the wall on a given mesh was found to depend strongly on the dissipation level.
11. Finally, it was shown in [7] that linear perturbations to lift or drag caused by numerical solutions containing point sources corresponding to stagnation pressure perturbations do appear to diverge towards the wall.

## 3 ANALYTIC ADJOINT SOLUTION FOR INCOMPRESSIBLE FLOW

Item 11 above hints at an adjoint singularity at the wall of the same nature as the well-known singularity along the incoming stagnation streamline discovered in [9]. An adjoint singularity at the wall would certainly explain the observed behavior of numerical solutions, but it would remain to determine the origin and characteristics of the singularity and to explain, likewise, how a singular (i.e. infinite) adjoint solution could be reconciled with the adjoint wall b.c. and sensitivity derivatives.

Fortunately, an analytic solution for the lift and drag-based adjoint two-dimensional incompressible Euler equations was obtained in [2] using the Green's function approach [10]. The resulting drag-based adjoint solution is smooth, but the lift-based adjoint solution contains singularities at the trailing edge, due to the sensitivity of the Kutta condition to perturbations of the flow, and also along the incoming stagnation streamline and the wall (see Figure 2 and Figure 3). Further, it can be shown that two particular linear combinations of adjoint variables, which yield the continuous adjoint sensitivity derivatives and the adjoint wall boundary conditions, respectively, are actually free of singularities.

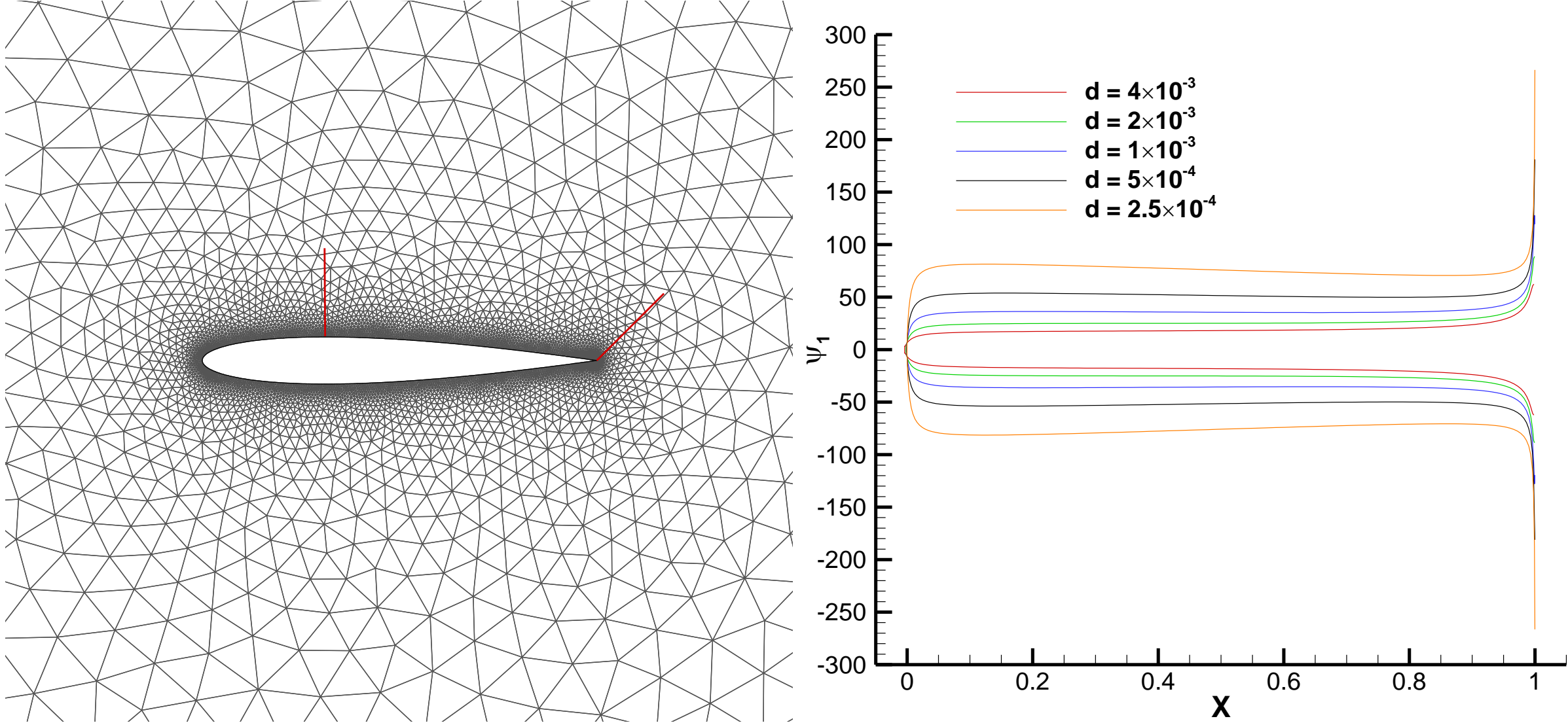

**Figure 2.** Left: detail of initial triangular mesh. Right: Lift-based adjoint variable $\boldsymbol{\psi}_1$ on a sequence of O-shaped curves surrounding the airfoil (right). The O-curves are built in the circle plane as circumferences concentric with the circle and are transferred to the airfoil plane with (1). d denotes distance to the wall on the circle plane.

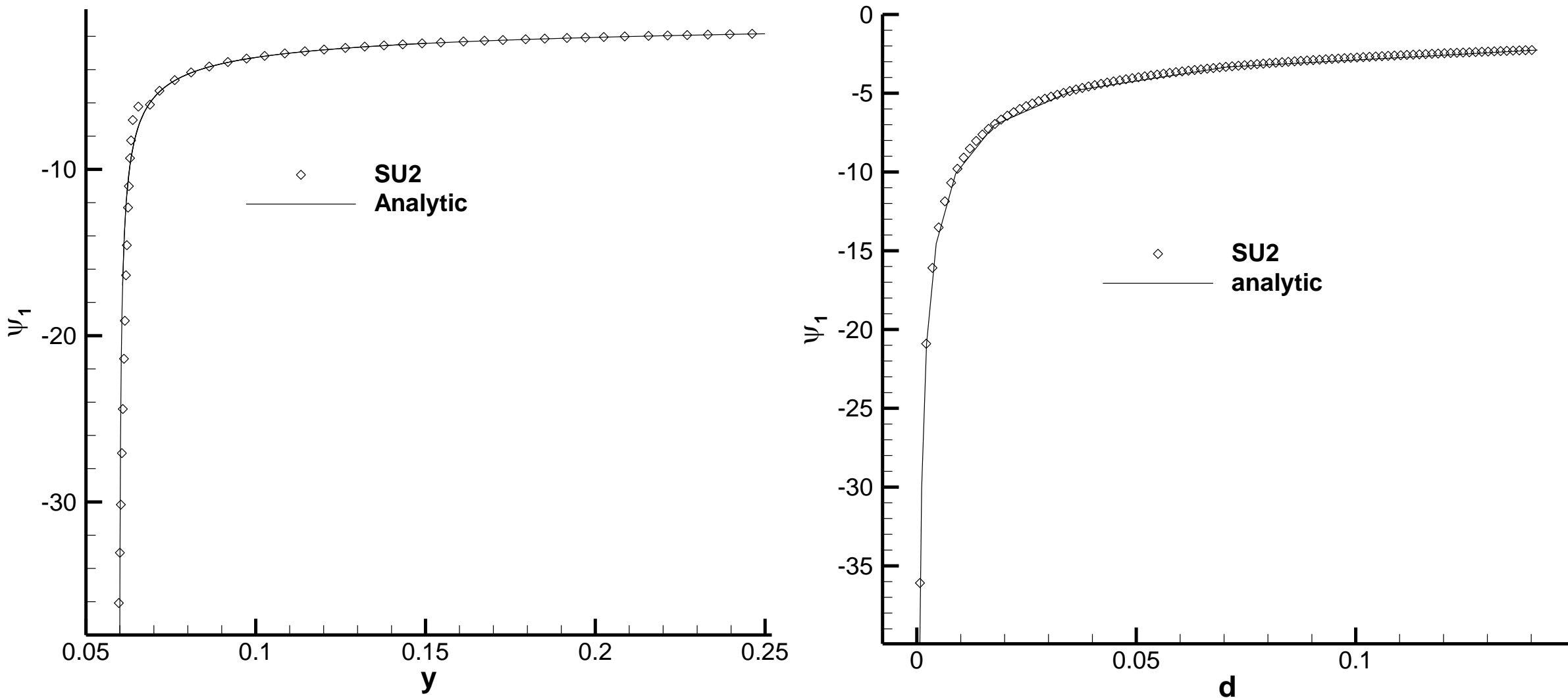

**Figure 3.** Lift-based adjoint variable $\boldsymbol{\psi}_1$ along a line normal to the airfoil wall at $x/c$ = 0.31 (left) and a line approaching the trailing edge (right) as indicated in Figure 2 (left). d denotes distance to the trailing edge.

## 4 CONCLUSIONS

The near-wall mesh divergence of solutions to the adjoint Euler equations occurring at subsonic and transonic speeds is reviewed. By examining a recently derived analytic adjoint solution, it is shown that the anomaly observed in numerical computations reflects the divergence of the analytic solution at the wall. On the numerical side, the numerical viscosity of the solver stabilizes the divergence, producing a finite value at the wall which nevertheless varies continually as the grid spacing or the intensity of the numerical dissipation change.

## ACKNOWLEDGEMENTS

The research described in this paper has been supported by INTA and the Ministry of Defence of Spain under the grants Termofluidodinámica (IGB99001) and IDATEC (IGB21001). The numerical computations reported in the paper have been carried out with the SU2 code, an open source platform developed and maintained by the SU2 Foundation.

## REFERENCES

[1] C. Lozano, "Watch Your Adjoints! Lack of Mesh Convergence in Inviscid Adjoint Solutions," *AIAA J.,* vol. 57, no. 9, pp. 3991-4006, 2019. https://doi.org/10.2514/1.J057259.

[2] C. Lozano and J. Ponsin, "Analytic Adjoint Solutions for the 2D Incompressible Euler Equations Using the Green's Function Approach," *Journal of Fluid Mechanics,* vol. 943, A22, 2022, doi:10.1017/jfm.2022.415.

[3] J. Katz and A. Plotkin, Low Speed Aerodynamics, 2nd edition, New York: Cambridge University Press, 2001.

[4] T. D. Economon, F. Palacios, S. R. Copeland, T. W. Lukaczyk and J. J. Alonso, "SU2: An Open-Source Suite for Multiphysics Simulation and Design," *AIAA Journal,* vol. 54, no. 3, pp. 828-846, 2016. doi: 10.2514/1.J053813.

[5] C. Lozano and J. Ponsin, "On the Mesh Divergence of Inviscid Adjoint Solutions," in *Proceedings of WCCM-ECCOMAS2020. F. Chinesta, R. Abgrall, O. Allix and M. Kaliske, (Eds). Scipedia, Volume 1300 - Inverse Problems, Optimization and Design. URL: https://www.scipedia.com/public/Lozano_Ponsin_2021a. DOI: 10.23967/wccm-eccomas.2020.258*, 2021.

[6] D. Schwamborn, T. Gerhold and R. Heinrich, "The DLR TAU-Code: Recent Applications in Research and Industry," in *ECCOMAS CFD 2006, European Conference on CFD*, Egmond aan Zee, The Netherlands, 2006.

[7] J. Peter, F. Renac and C. Labbé, "Analysis of finite-volume discrete adjoint fields for two-dimensional compressible Euler flows," *Journal of Computational Physics,* vol. 449, p. 110811, 2022. Doi: 10.1016/j.jcp.2021.110811.

[8] D. Ekelschot, "Mesh adaptation strategies for compressible flows using a high-order spectral/hp element discretisation," Ph.D. Thesis, Department of Aeronautics, Imperial College (London), 2016.

[9] M. B. Giles and N. A. Pierce, "Adjoint Equations in CFD: Duality, Boundary Conditions and Solution Behavior," AIAA Paper 97–1850, 1997. doi: 10.2514/6.1997-1850.

[10] M. B. Giles and N. A. Pierce, "Analytic adjoint solutions for the quasi-one-dimensional Euler equations," *J. Fluid Mechanics,* vol. 426, pp. 327-345, 2001. doi: 10.1017/S0022112000002366.